\documentclass[aoas,preprint]{imsart}

\RequirePackage{natbib}
\RequirePackage[OT1]{fontenc}
\RequirePackage{amsthm,amsmath}
\RequirePackage[colorlinks,citecolor=blue,urlcolor=blue]{hyperref}
\usepackage{amssymb,graphicx,lscape}
\startlocaldefs
\numberwithin{equation}{section}
\theoremstyle{plain}
\newcommand{\m}[1]{\mbox{\bf{#1}} }
\endlocaldefs

\begin{document}

\begin{frontmatter}
\title{Joint Mean and Covariance Modeling of Multiple Health Outcome Measures}
\runtitle{Joint Mean and Covariance}
%\thankstext{T1}{}

\begin{aug}
\author{\fnms{Xiaoyue} \snm{Niu}\thanksref{t2,m1}}
\and
\author{\fnms{Peter D.} \snm{Hoff}\thanksref{t3,m2}}

\thankstext{t2}{Supported by NIH grant R01 AI36664-01}
\thankstext{t3}{Supported by NSF grant DMS-1505136}
\runauthor{X. Niu and P. D. Hoff}

\affiliation{The Pennsylvania State University\thanksmark{m1} and Duke University\thanksmark{m2}}

\end{aug}

\begin{abstract}
Health exams determine a patient's health status by comparing the patient's measurement with a population reference range, a 95\% interval derived from a homogeneous reference population. Similarly, most of the established relation among health problems are assumed to hold for the entire population. We use data from the 2009 {\textendash} 2010 National Health and Nutrition Examination Survey (NHANES) on four major health problems in the U.S. and apply a joint mean and covariance model to study how the reference ranges and associations of those health outcomes could vary among subpopulations. We discuss guidelines for model selection and evaluation, using standard criteria such as AIC in conjunction with posterior predictive checks. The results from the proposed model can help identify subpopulations in which more data need to be collected to refine the reference range and to study the specific associations among those health problems. 
\end{abstract}


\end{frontmatter}

\section{Introduction}

Health exams, a patient's health status by comparing the patient's measurement with a population reference range. For example, in measuring blood sugar levels, the normal range of a fasting glucose level is 70 to 100 mg/dl.  People with values lower than 70 mg/dl are considered to have hypoglycemia (low blood sugar), and people with values higher than 100 mg/dl are considered pre-diabetic (100 {\textendash} 125 mg/dl) or diabetic ($>$ 125 mg/dl). The reference range is usually a 95\% interval  derived from a reference population. Current guidelines suggest that if the reference population is heterogeneous, we should partition it and provide a separate reference range for each subpopulation (\cite{CLSI2008}). \cite{mattix02} argue that using a single cutpoint in diagnosing kidney disease for both genders and various race groups biases the prevalence of the disease for some subpopulations and thus underestimates their risks. The most widely used partition guideline is that if the ratio of the two subpopulation standard deviations is greater than 1.5, we should collect large enough samples in those groups and provide separate reference ranges, regardless of whether the mean difference is significant or not(\cite{harris90}). 

Similarly, certain health problems are associated with others. For example, obesity is a risk factor of diabetes. Most established relations among health problems are assumed to hold for the entire population. However, those relations could vary among subpopulations. For example, \cite{foulds12} find that the relation between obesity and diabetes varies with ethnicity. 
 
We use data from the 2009 {\textendash} 2010 National Health and Nutrition Examination Survey (NHANES) to look at some major health problems in the U.S. (\cite{nhanes_2010}, \url{http://www.cdc.gov/nchs/nhanes/search/nhanes09_10.aspx}). NHANES is designed to assess the health and nutritional status of adults and children in the United States. It collects participants' demographic, socio-economic, dietary, activity, and behavioral  information through interviews in their homes. It also performs physical measurements and blood and urine tests in mobile examination centers. The National Center for Health Statistics (NCHS),  part of the Centers for Disease Control and Prevention (CDC), conducts the survey mainly to determine the prevalence of major diseases and risk factors in the U.S. population.  

We focus on four health problems that are believed to be associated: chronic kidney disease (CKD), obesity, hypertension, and diabetes. The severity and progression of each health problem can be assessed by a quantitative measurement. In chronic kidney disease, defined as abnormalities of kidney structure or function, the kidneys are damaged and cannot filter blood as needed. Kidney damage and disease progression can be assessed by the urine albumin/creatinine ratio (ACR). ACR below 30 mg/g is considered normal and above 30 mg/g is considered to indicate microalbuminuria, a marker for CKD and kidney damage (\cite{kdigo12}). Obesity is quantified by body mass index (BMI), defined as weight in kilograms divided by the square of height in meters. For adults of 20 years and older, a BMI below 18.5 is considered underweight, 18.5 {\textendash} 24.9 is normal, 25 {\textendash} 29.9 is overweight, and over 30 is obese. Hypertension (high blood pressure) is diagnosed by measuring both systolic blood pressure (SBP) and diastolic blood pressure (DBP). A normal blood pressure corresponds to SBP of 80 {\textendash} 120 mmHg AND DBP of 60 {\textendash} 80 mmHg. If SBP $>$ 120 or DBP $>$ 80, a patient is considered to have elevated blood pressure. If SBP $>$ 120 and DBP $>$ 80, a patient is considered to have hypertension. If SBP $<$ 80 or DBP $<$ 60, a patient is considered to have hypotension (low blood pressure). SBP and DBP are usually correlated, so we take DBP to represent blood pressure (BP). Finally, a common measurement for diabetes is the fasting glucose level (GLU), discussed earlier. 

All of these reference ranges are derived by assuming that the measurement comes from a homogeneous population, summarized by its mean and variance. Based on the CLSI guidelines and some previous findings (\cite{ckd_prev}, \cite{fraser_2012}), we use gender, age, race/ethnicity, and education level to to define subpopulations. Refining the reference ranges and associations of the four health problems requires estimation of the means and covariances of ACR, BMI, BP, and GLU in the subpopulations. To estimate how the mean and covariance structure vary among subpopulations, we jointly model them as functions of the demographic variables. 

Joint regression models for means and covariances have been developed mainly in the context of  longitudinal and repeated-measures studies. \cite{liang_1986} and \cite{zeger_1986} use generalized estimating equations (GEE) to simultaneously estimate the parameters in the mean and covariance of a longitudinal response vector, which improves the efficiency of the mean estimate substantially. When the heteroscedasticity is temporal, multivariate autoregressive conditionally heteroscedastic (ARCH) models are well-studied in the econometric literature (\cite{engle_kroner_1995},  \cite{fong_2006}). The approach proposed by \cite{pourahmadi_1999} uses the Cholesky decomposition to parameterize the class of positive-definite covariance matrices by expressing the unconstrained parameters through generalized linear models. However, this model is not invariant to reorderings of the response, and thus might not be appropriate for studies without longitudinal or spatial structure.  \cite{chiu_1996} model the logarithm of the covariance matrix as linear functions of the explanatory variables, although the parameters are somewhat difficult to interpret. \cite{pourahmadi_2011} gives a comprehensive literature review of covariance estimation models. 

\cite{hoff_2012} propose a covariance regression model that directly models the covariance matrix as a function of explanatory variables. In this natural extension of the mean regression model, the parameters have interpretations similar to those in a mean regression. However, \cite{hoff_2012} focus on model development and geometric interpretations. They discuss an example of a single continuous predictor. We extend the covariance regression model of \cite{hoff_2012} to accommodate multiple categorical predictor variables. We also discuss practical issues with real data, including model selection and how to present and interpret the results. 

In the next section, we explore some basic features of the NHANES data. In Section 3 we introduce the proposed method for joint modeling and outline the process of model selection. We describe the details of model selection and present our main findings in Section 4. Discussion follows in Section 5. 

\section{The NHANES Data} \label{sec:eda}
The 2009 {\textendash} 2010 NHANES had 10,537 participants, but only 3,386 had a fasting glucose blood test.  Among them, only 2,613 (77\%) had data for all four of ACR, BMI, BP (DBP), and GLU. Because GLU is an important measurement when dealing with kidney diseases, we use in this analysis only those who have complete data; the sample size reduction is due mainly to the small number of participants who took the blood tests (GLU). Further discussion of the sample size and missing data is in Section 5. 

The demographic variables are categorical: gender (male, female), age (20 {\textendash} 39, 40 {\textendash} 59, 60 {\textendash} 79, 80+), race/ethnicity (in order of decreasing sample size: non-Hispanic white, Mexican American, non-Hispanic black, other Hispanic, and other), education (less than $9^{th}$ grade, $9^{th}$ to $11^{th}$ grade, high school, associate degree or some college, and college degree and higher). For the 16 marginal groups defined by one category of one predictor, such as male, the sample sizes all exceed 100. The sample sizes among the 93 two-way cells range from 4 to 660 (median 132). Of the 200 four-way cells, 28 are empty, and the median sample size among the other 172 four-way cells is 7.5 (range 1 {\textendash} 74; quartiles 2 and 17). A detailed sample size tabulation is in Supplemental Material A (\cite{niu18}).

\begin{figure}[!ht]
\begin{center}
\includegraphics[height=4.5in]{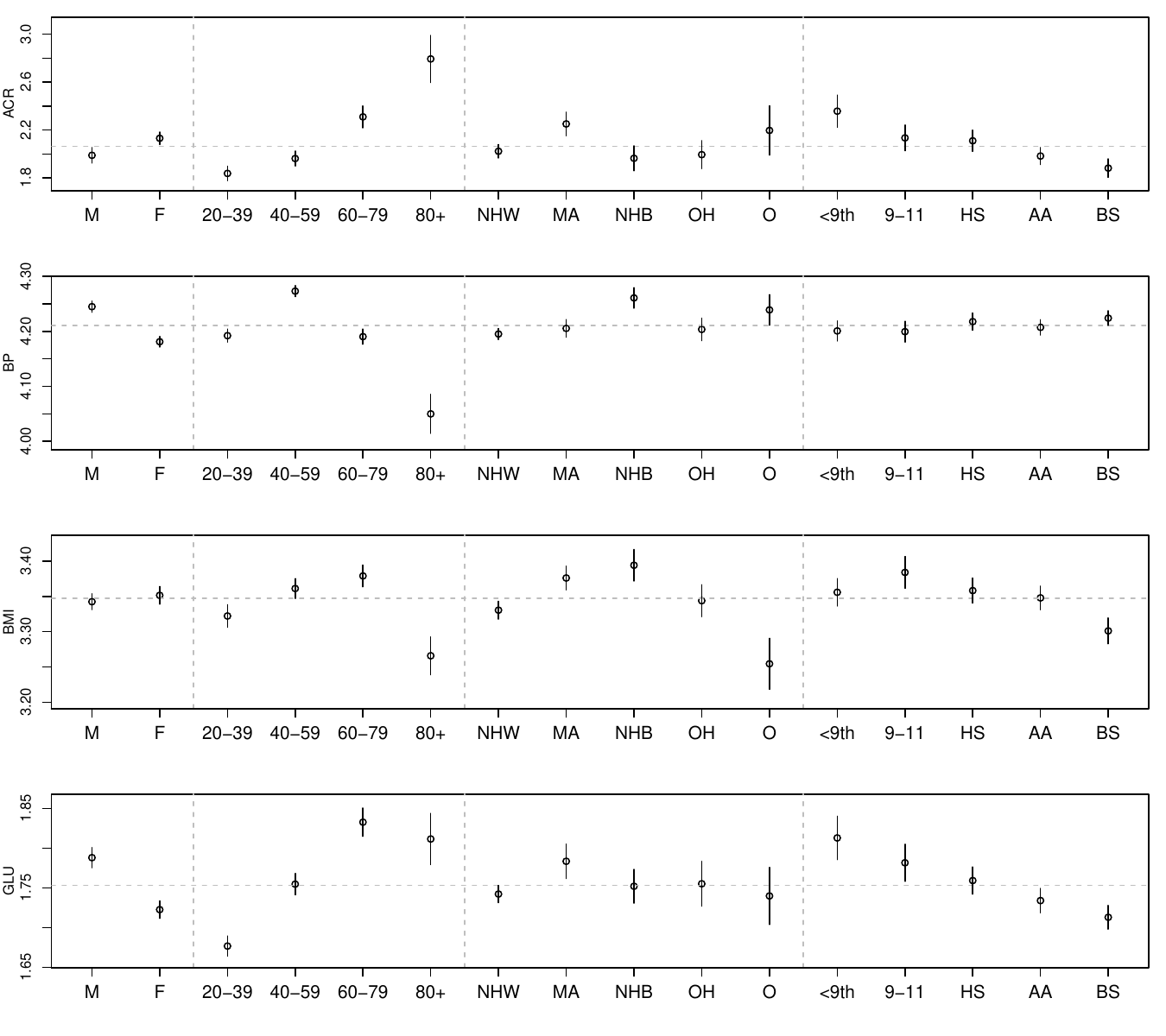}
\end{center}
\caption{Sample means with 95\% confidence intervals of albumin creatinine ratio (ACR), diastolic blood pressure (BP), body mass index (BMI), and glucose (GLU), on the natural log scale, by gender, age, race/ethnicity (NHW: non-Hispanic white, MA: Mexican American, NHB: non-Hispanic black, OH: other Hispanic, O: other), and education ($<$ 9th: less than $9^{th}$ grade, 9 {\textendash} 11: $9^{th}$ to $11^{th}$ grade, HS: high school, AA: associate degree or some college, BS: college degree and higher) categories. The horizontal dotted line indicates the pooled sample mean. The figure is based on the subset of 2613 individuals from NHANES 2009 {\textendash} 2010 who have complete observations on these four variables.} \label{fig:mean}
\end{figure}

%\clearpage

\begin{figure}[!ht]
\begin{center}
\includegraphics[height=4.5in]{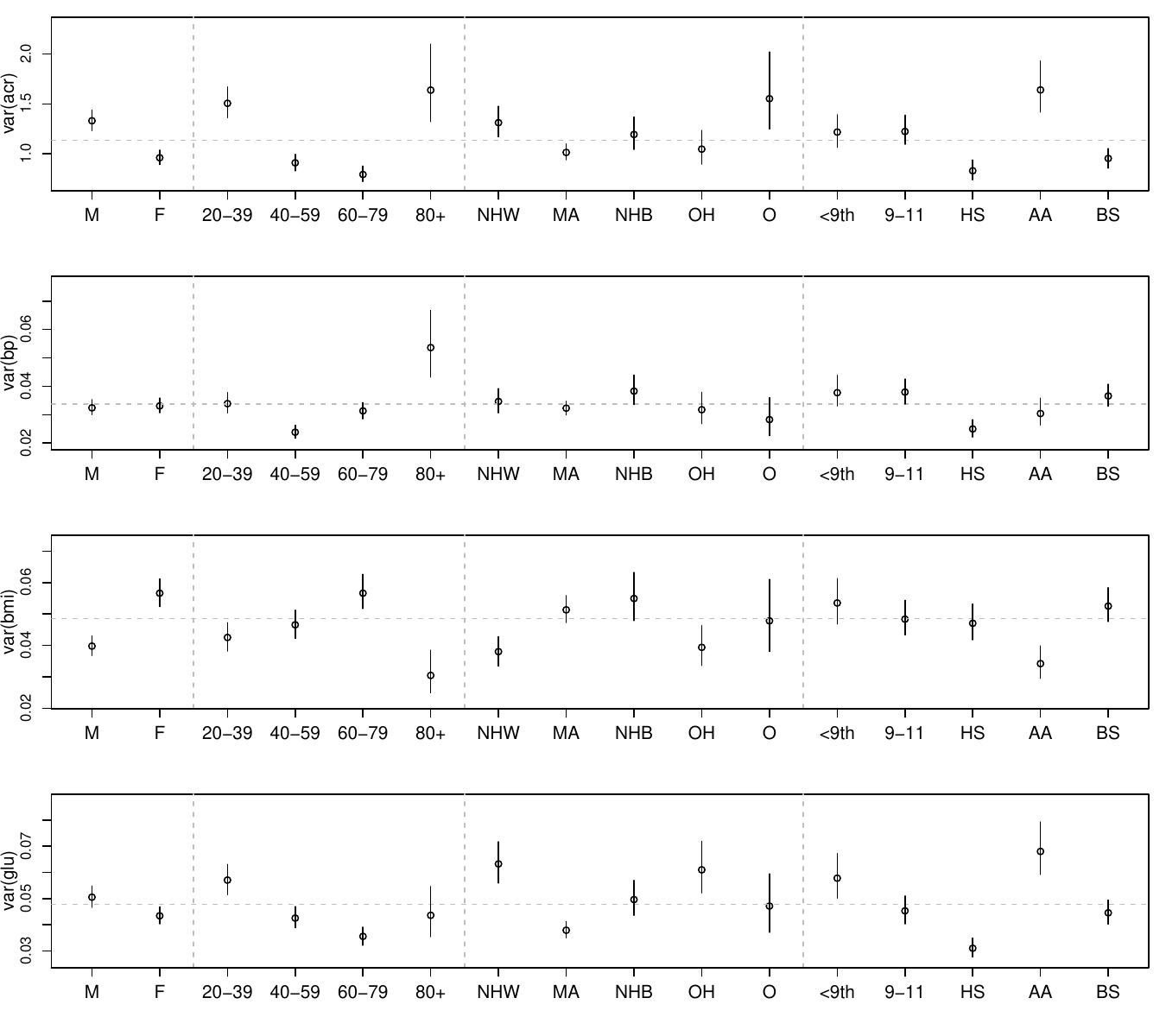}
\end{center}
\caption{Sample variances with 95\% Bayesian posterior intervals of albumin creatinine ratio (ACR), diastolic blood pressure (BP), body mass index (BMI), and glucose (GLU), on the natural log scale, by gender, age, race/ethnicity, and education categories. The horizontal dotted line indicates the pooled sample variance. The figure is based on the subset of 2613 individuals from NHANES 2009 {\textendash} 2010 who have complete observations on these four variables. } \label{fig:var}
\end{figure}

%\clearpage

\begin{figure}
\hspace{-0.5cm}
%\begin{center}
\includegraphics[height=5in]{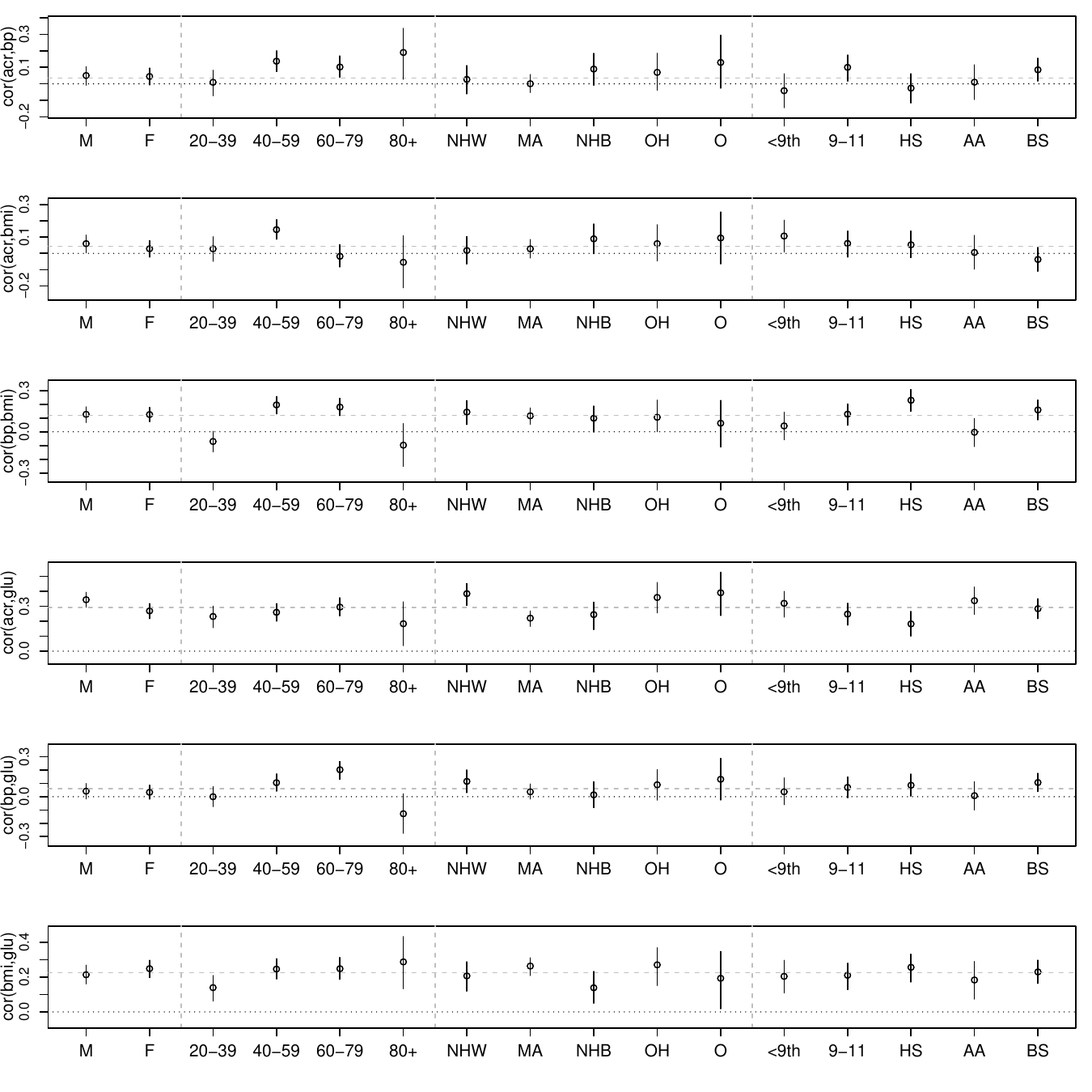}
%\end{center}
\caption{Sample pairwise correlations, with 95\% Bayesian posterior intervals, for each pair of albumin creatinine ratio (ACR), diastolic blood pressure (BP), body mass index (BMI), and glucose (GLU), on the natural log scale, by gender, age, race/ethnicity, and education categories. The horizontal dotted line is at 0. The horizontal dashed line indicates the all-sample correlation. The figure is based on the subset of 2613 individuals from NHANES 2009 {\textendash} 2010 who have complete observations on these four variables. } \label{fig:cor}
\end{figure}

%\clearpage

Because most of them are skewed, we analyze these variables on the natural log scale. In Figure \ref{fig:mean}, the sample means of the health measurements vary greatly among demographic groups. We also calculate the sample covariance matrices within the demographic groups, along with Bayesian posterior intervals using a non-informative Wishart prior. Figure \ref{fig:var} and Figure \ref{fig:cor} show that the variances and correlations also vary among subpopulations. For example, the variance of ACR is about 1.14 overall, but it can be as low as 0.79 in the 60 {\textendash} 79 age group and as high as 1.64 in the 80+ age group. Similarly, the correlation between BP and GLU is around 0.06 overall, but it can be as low as -0.13 in the 60 {\textendash} 79 age group and as high as 0.2 as in the 80+ age group. We have also considered multiplicity of the intervals. Supplemental Material B (\cite{niu18}) includes Bonferroni-corrected simultaneous 95\% intervals. The patterns are very similar to Figures \ref{fig:mean} to \ref{fig:cor}. The exploratory findings suggest the need for a statistical analysis that allows both the mean and covariance matrix of these health outcomes to vary among demographic groups.

\section{Statistical models}
\subsection{The covariance regression model} \label{sec:model}
Our goal is to describe and estimate heterogeneity of means and covariances for the cross-classified groups defined by gender, age, race/ethnicity, and education. Specifically, let $\m y$ be the 4-dimensional vector of the logarithms of ACR, BP, BMI, and GLU of an individual, and let $\m x$ be a covariate vector describing the individual's gender, age, race/ethnicity, and education level. We would like to estimate $\rm E[\m y|\m x]=\boldsymbol  \mu_{\m x}$ and $\rm {Cov}[\m y|\m x]=\boldsymbol \Sigma_{\m x}$ simultaneously. 

The small number of observations for each combination of categories makes it impractical to estimate a separate covariance matrix $\boldsymbol \Sigma_{\m x}$ for each group, based on data from only that group. On the other hand, a common covariance matrix for all groups would misrepresent the relations among the response variables and result in loss of efficiency for the mean parameters (see \cite{glm}, Chap 9 and Chap 10). More-flexible models share information across covariance matrices. \cite{boik_2002, boik_2003} assumes common principal components of the covariance matrix. \cite{hoff_2009} proposes shrinking the covariance matrix toward a common eigenvector structure with varying degrees of shrinkage among principal components. \cite{cripps_2005} show efficiency improvement for the mean regression parameters with a carefully selected covariance model. \cite{gaskins_2013} give a more-comprehensive review of pooling methods.

Besides avoiding the loss of efficiency from assuming a common covariance structure, we are also interested in how $\boldsymbol \Sigma_{\m x}$ varies with $\m x$. An alternative way of pooling across groups addresses this question with a covariance regression model that  parsimoniously describes heteroscedasticity
among groups (\cite{hoff_2012}). The proposed model parametrizes the mean and covariance of a multivariate response vector as  parsimonious functions of explanatory variables. This approach allows joint modeling of the mean and covariance structure of the population being studied.

To introduce the model, we adopt the notation and definitions in \cite{hoff_2012}. Let $\m y \in {\mathbb R}^p$ be a random multivariate response vector
and $\m x_1\in {\mathbb R}^{q_1}$ and $\m x_2 \in {\mathbb R}^{q_2}$ be vectors of explanatory variables. The variables in $\m x_1$ and $\m x_2$ can overlap or even be the same. Denote the mean of $\m y|\m x$ as $\boldsymbol  \mu_{ \m x}=\rm E[\m y|\m x]$ and the $p\times p$ covariance matrix of $\m y|\m x$ as $\boldsymbol \Sigma_{ \m x} = \rm {Cov}[\m y|\m x]$.
The covariance regression model has the form
\begin{eqnarray}
\boldsymbol  \mu_{\m x_1}&=&\m B_1 \m x_1, \label{eqn:mean}\\
\boldsymbol  \Sigma_{\m x_2} &=& \m A + \m B_2 \m x_2\m x_2^T \m B_2^T,
\label{eqn:cov1}
\end{eqnarray}
where $\m B_1$ is a $p \times q_1$ matrix, $\m A$ is a $p\times p$ positive-definite matrix, and $\m B_2$ is a
$p\times q_2$ matrix. The resulting covariance function in Equation (\ref{eqn:cov1}) is positive-definite for
all $\m x_2$, and it expresses the covariance as a constant covariance
matrix $\m A$ plus a rank-1, positive-semi-definite matrix that varies with $\m x_2$. \cite{hoff_2012} consider the case where $\m x_2$ is a single continuous variable, which makes the variance a quadratic function of the predictor. We apply such a model when there are multiple categorical predictors. For example, if sex is our only predictor, we let $\m x_2= (1,1)^T $ for males and $\m x_2=(1,0)^T$ for females, where the first ``1'' in each vector corresponds to the intercept.

The covariance regression model can also be interpreted as a special random-effects model.
Assume the observed data $\m y_1,\ldots, \m y_n$ are generated by the following model:
\begin{eqnarray}
\m y_i &=& \boldsymbol  \mu_{{\m x}_{1i}}  + \gamma_i \times \m B_2 \m x_{2i} + \boldsymbol  \epsilon_i \label{eqn:re}\\
\mbox{E}[ \boldsymbol \epsilon_i ]& =& \m 0  \ , \  \mbox {Cov}[\boldsymbol \epsilon_i]\ = \ \m A \nonumber \\
\mbox{E}[ \gamma_i ]& =&  0  \ \, , \   \mbox {Var}[\gamma_i]  \ = \ 1  \  , \
\mbox{E}[\gamma_i\times  \boldsymbol \epsilon_i ]  \  = \ \m 0 .
\nonumber
%\E{\gamma_i\times  \boldsymbol \epsilon_i } &=&  \nonumber 0
\end{eqnarray}
We can interpret $\gamma_i$ as describing
additional individual-level variability beyond the random error $\boldsymbol  \epsilon_i$.
The row vectors $\{\boldsymbol  b_{21},\ldots, \boldsymbol  b_{2p}\}$ of the coefficient matrix $\m B_2$ describe how this
additional variability is manifested in
the $p$ response variables.

Model (\ref{eqn:cov1}) restricts the difference between $\boldsymbol \Sigma_{ \m x}$
and the constant matrix $\m A$ to be a rank-1 matrix. This rank-1 model essentially requires that the residuals of the $p$ responses are along the same direction. This restriction can be relaxed by allowing the difference from the constant covariance to have higher rank. For example, a rank-2 covariance regression model has the following form:
\begin{equation}
\m y_i=\boldsymbol  \mu_{\m x_{1i}} + \gamma_i\times \m B_2 \m x_{2i}+ \psi_i \times \m B_3 \m x_{2i}+\boldsymbol  \epsilon_i ,
\label{eqn:re2}
\end{equation}
where $\gamma_i$ and $\psi_i$ are  mean-zero variance-one random variables,
uncorrelated with each other and with $\boldsymbol  \epsilon_i$. $\m B_2$ in Equation (\ref{eqn:re2}) has a different estimate and interpretation than the $\m B_2$ in Equation (\ref{eqn:re}) because of the additional term in Equation (\ref{eqn:re2}). We keep the same notation for simplicity.
Under this model, the covariance matrix of $\m y_i$ is given by
\begin{equation}
\boldsymbol \Sigma_{\m x_2} =\m A  + \m B_2\m x_2 \m x_2^T \m B_2^T + \m B_3\m x_2 \m x_2^T \m B_3^T.
\label{eqn:cov2}
\end{equation}
Model (\ref{eqn:cov2}) allows the deviation of $\boldsymbol \Sigma_{\rm x_2}$ from the constant matrix 
$\m A$ to have rank 2. We can interpret the second random effect
$\psi$ in Equation (\ref{eqn:re2}) as allowing an additional, independent source of heteroscedasticity for
the $p$ response variables.
Further flexibility can be gained with additional random effects,
allowing the difference between $\boldsymbol \Sigma_{ \m x}$ and the constant matrix $\m A$
to be of any desired  rank up to $p$.

Assuming normality of the error terms, the rank-1 model can be expressed as follows:
\begin{eqnarray}
\gamma_1,\ldots, \gamma_n & \stackrel{\rm iid}{\sim} & \mbox{normal}(0,1) \nonumber \\
\boldsymbol \epsilon _1,\ldots, \boldsymbol \epsilon_n & \stackrel{\rm iid}{\sim} & \mbox{multivariate normal}(\m 0,\m A)  \label{eqn:nre}  \\
\m y_i & =& \boldsymbol  \mu_{\m x_{1i}} + \gamma_i \times \m B_2\m x_{2i} + \boldsymbol  \epsilon_{i} \nonumber \\
\boldsymbol \mu_{\m x_{1i}} &=& \m B_1\m x_{1i}. \nonumber
\end{eqnarray}

Parameters of this normal covariance regression model can be estimated by maximum likelihood via the EM algorithm or by Bayesian estimation via Markov chain Monte Carlo (MCMC). We focus on Bayesian estimation mainly for three reasons: 1. the convergence of EM can be very slow due to the identifiability issue; 2. it is easier to obtain the intervals for those identifiable parameters in the Bayesian setting; and 3. it is easier to perform model selection and diagnoses (such as posterior predictive checks discussed below). Calculations are facilitated by using a semi-conjugate prior
distribution for $\m A, \m B_1$ and $\m B_2$, in which
 $p(\m A)$ is an inverse-Wishart$(\m A_0^{-1},\nu_0)$ distribution and $\m C=(\m B_1, \m B_2)$ has a matrix normal distribution.

\cite{hoff_2012} introduce this covariance regression model in the context of continuous predictors, and give an example with one continuous predictor. We extend that model to jointly estimate the mean and covariance structure of a large number of groups defined by several categorical variables. We fit the model in Equation (\ref{eqn:mean}) and (\ref{eqn:cov1}) and its higher-rank version to the NHANES data, allowing different sets of predictors for the mean and covariance matrix. Including multiple categorical predictors requires that we address practical issues such as variable selection and evaluation, which are not discussed in \cite{hoff_2012}. In the next section, we discuss the outline of model selection for this covariance regression model with multiple categorical factors.

\subsection{Model selection and evaluation} \label{sec:selection}
Similar to any regression model, the covariance regression model requires a procedure for variable selection. The process has three components: mean variable selection, covariance variable selection, and covariance rank selection. As noted in \cite{hoff_2012}, because of the non-identifiability of some of the parameters in the higher-rank model, methods such as AIC or BIC are not directly applicable when comparing models with different ranks. For the NHANES data, we include 4 predictors with at most 2-way interactions (6 interaction terms) in both the mean and covariance models. The maximum possible rank for a 4-dimensional response is 4. Simultaneous selection of the appropriate interaction terms for the meanmodel and the covariance model and the selection of rank would require $2^6 \times 2^6 \times 4$ evaluations of the model, which is computationally impractical. 

As an alternative, we propose a ``forward search procedure'' that tries to find the most parsimonious model without obvious lack of fit. We outline the procedure in this section and elaborate the details with data in Section 4. First, we simplify the situation by separating the tasks of mean and covariance model selection, based on the fact that under multivariate normality, the maximum likelihood estimator of the mean parameters is consistent under mis-specification of the covariance structure (\cite{cox_1987}). For selecting the mean model, we assume a homogeneous covariance model and use a standard variable-selection criterion such as AIC or BIC. Next we fix that mean model and fit the simplest covariance model, a rank-1 model with only main effects of the four predictors. Then we assess goodness of fit. If the simplest model has only moderate lack of fit, we add one interaction term at a time until we find a model that is acceptable. If the simplest model displays serious lack of fit, we increase the rank by 1 and implement forward selection for the rank-2 model until we find an acceptable set of predictors. If necessary, we can continue the selection to rank 3 or higher. 

Model fit is evaluated with posterior predictive distributions (\cite{guttman67} and \cite{rubin84}). To assess the model, we need to construct a meaningful statistic to represent lack of fit. We would like to make sure that the model we select generates predictive datasets $\tilde {\m{Y}}$ that resemble the observed dataset $\m Y$ (the observed response matrix) in terms of features that are of interest. The key idea is that the population is not homogeneous, and the covariance matrices differ among the subpopulations defined by the variables that make up $\m x_2$. Therefore we construct a diagnostic statistic that describes the heteroscedasticity across subpopulations defined by pairwise combinations of the factors, such as all the white females, or people 30 {\textendash} 49 years old with a high school degree. We use 2 variables instead of 4 because the sample size is not large enough for estimation of all groups obtained by cross-classifying the 4 variables. We define the posterior check statistic 
\begin{equation*}
t_{hk}(\m Y)=\sum_{x_h,x_k}[\mbox{tr}(\m S_0^{-1} \m S_{x_h,x_k})-\log|\m S_0^{-1} \m S_{x_h,x_k}|],
\end{equation*}
where $\m S_0$ is the all-sample covariance matrix, $h$ and $k$ denote the factors (e.g., gender and age), $x_h$ and $x_k$ represent the levels of factors $h$ and $k$ (e.g., male and 20 {\textendash} 39 years old), and $\m S_{x_h,x_k}$ is the sample covariance matrix of the subpopulation defined by $x_h$ and $x_k$ (e.g., males 20 {\textendash} 39 years old). The statistic $t_{hk}(\m Y)$ is the sum of the Wishart kernels of the sample covariance matrices $\m S_{x_h,x_k}$ for all possible values of $x_h$ and $x_k$, with $\m S_0$ as the center. If the population is homogeneous, $\m S_0$ should be a good estimate of the $\m S_{x_h,x_k}$. The statistic $t_{hk}(\m Y)$ describes the discrepancy between $\m S_0$ and the $\m S_{x_h,x_k}$ and represents the heterogeneity of the subpopulation covariance matrices. For each model, we compute the posterior predictive distribution of $t_{hk}$ for all pairs of factors. We then compare the observed value with the posterior predictive distribution of $t_{hk}$. If the observed statistic lies in the tail of the posterior predictive distribution, it indicates lack of fit in that pair of factors. 

\section{Analysis of the NHANES data}
In this section, we first describe the detailed model-selection procedure and selection results for the NHANES data. Then we present the analysis results of the NHANES data using the covariance regression model.

\subsection{Model selection for the NHANES data}

\begin{figure}[!ht]
\begin{center}
\includegraphics[height=4.5in]{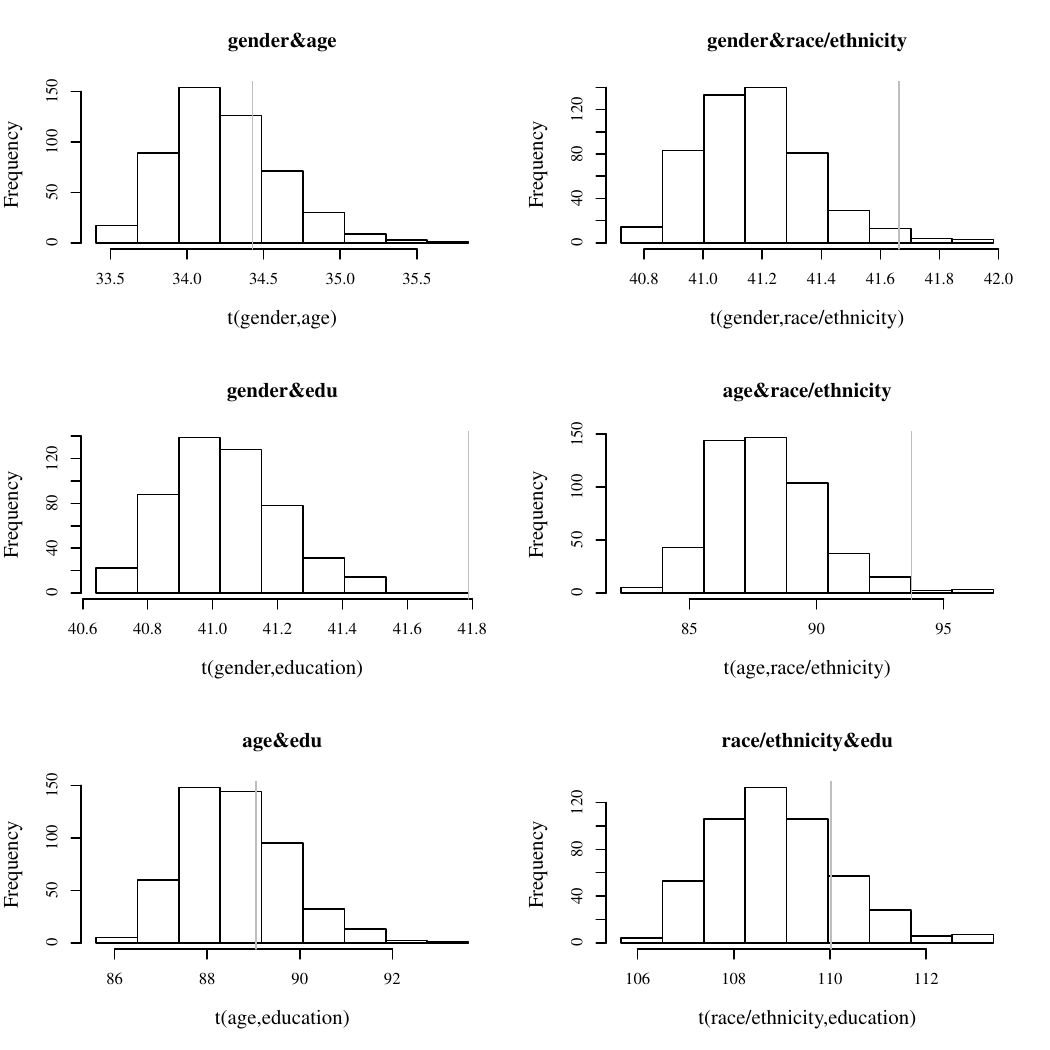}
\end{center}
\caption[pcheck12]{Posterior predictive distributions (of 800 posterior samples) of the rank-2 model with only main effects. In each histogram the vertical line represents the goodness-of-fit statistic calculated from the data.} \label{fig:m1r2}
\end{figure}

\begin{figure}[!ht]
\begin{center}
\includegraphics[height=4.5in]{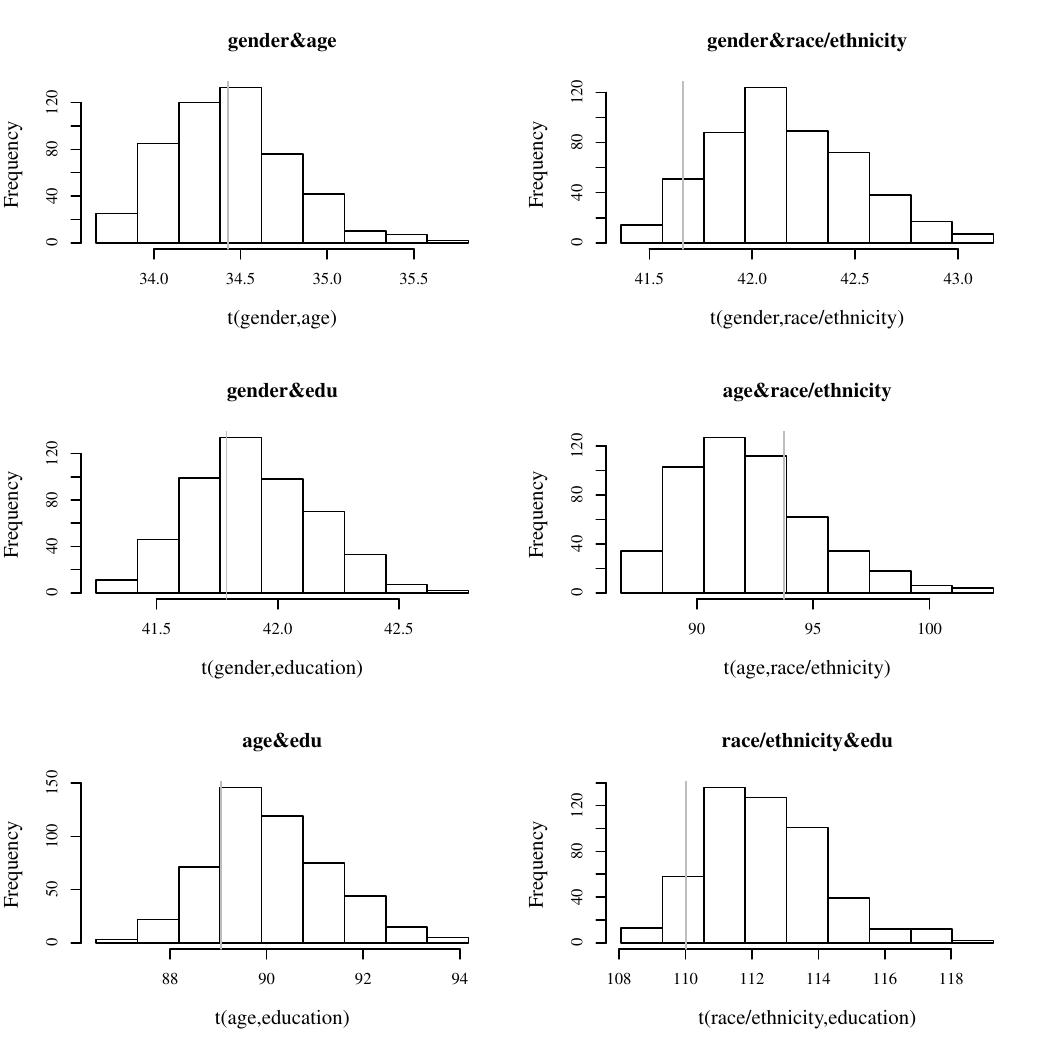}
\end{center}
\caption[pcheck12]{Posterior predictive distributions (of 800 posterior samples) of the rank-2 model with main effects and gender*(race/ethnicity) interaction. In each histogram the vertical line indicates the goodness-of-fit statistic calculated from the data.} \label{fig:m1r2t1}
\end{figure}

Following the outline in Section \ref{sec:selection}, we first assume a constant covariance model and use AIC (\cite{aic73}) to select the set of predictors for the mean model. The best mean model under AIC includes the main effects of gender, age, race/ethnicity, and education, as well as four two-way interactions: gender and age, gender and race/ethnicity, gender and education, and age and race/ethnicity, with a total of 36 parameters. We then fix the mean model and select the explanatory variables in the covariance model. We first fit a rank-1 covariance regression model with main effects of the four predictors. We examine goodness of fit of this model and find that the observed values of the test statistics lie in the tails of the posterior predictive distributions for all groups. We then fit a rank-2 model with main effects of the four predictors. We plot the posterior predictive distributions of the test statistics in Figure \ref{fig:m1r2}. Three of the six subpopulations (gender and age, age and education, and race/ethnicity and education) are well represented by the model. The remaining three subpopulations, which show lack of fit, all involve gender and/or race/ethnicity. Therefore, we add the interaction between gender and race/ethnicity to the rank-2 covariance model and present the goodness-of-fit diagnostics in Figure \ref{fig:m1r2t1}. The new model generally improves the goodness of fit from the rank-2 main-effects model, and it appropriately captures the heterogeneity in most of the 2-variable subpopulations. This relatively parsimonious model has no obvious lack of fit. We have also compared this model with adding other interaction terms and with a rank-3 model. None of those alternatives outperform this one. Therefore, our final model as stated in Equation (\ref{eqn:cov2}) and (\ref{eqn:nre}) includes the following terms in the mean and covariance structure:
\begin{eqnarray}
\boldsymbol \mu_{\m x_1} & \sim & \mbox{GENDER+AGE+RACE/ETHNICITY+EDU} \nonumber \\
      &+& \mbox{GENDER*AGE+GENDER*(RACE/ETHNICITY)}\nonumber \\
			&+& \mbox{GENDER*EDU+AGE*(RACE/ETHNICITY)}, \nonumber \\
\boldsymbol \Sigma_{\m x_2} & \sim & \mbox{GENDER+AGE+RACE/ETHNICITY+EDU} \nonumber\\
&+& \mbox{GENDER*(RACE/ETHNICITY)}.
\label{eqn:model_final}
\end{eqnarray}

%\clearpage

\subsection{Results for NHANES data}

We fit the final model in Equation (\ref{eqn:model_final}) and obtain the Bayesian estimates through Gibbs sampling, using the priors described in Section \ref{sec:model}. We run an MCMC chain for 50,000 iterations with thinning of every 50 samples (i.e. use every 50th iteration), drop the first 200 post-thinning samples as burn-in, and check the trace plots of key quantities for convergence. The analysis is performed using R-2.15.1, package ``covreg''. The key code to fit the model and summarize the results is in Supplemental Material C (\cite{niu18}). The computation time is about 6 hours on a PC with an i5 core. It remains as future work to speed up the package in order to handle larger datasets. 

\begin{figure}[!ht]
\begin{center}
\includegraphics[height=3in]{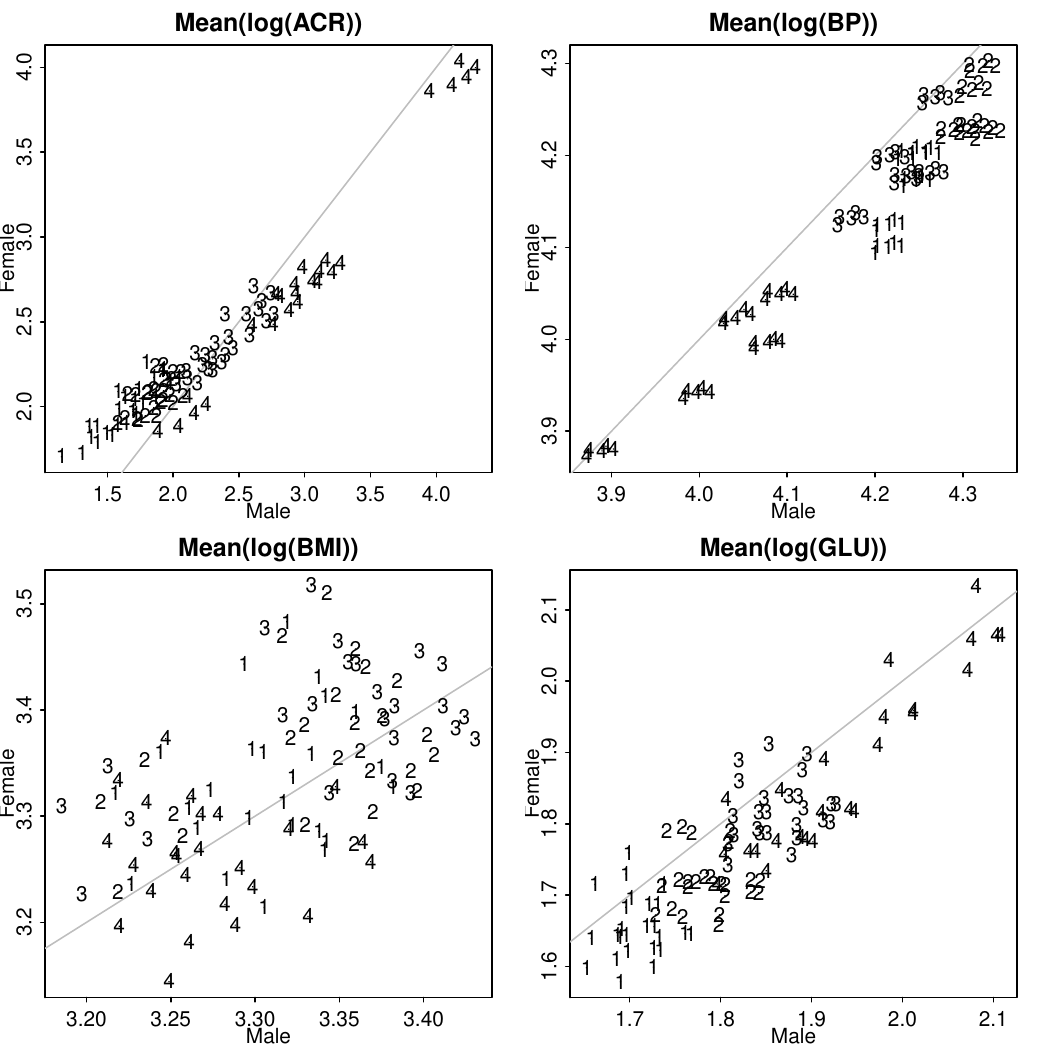}
\end{center}
\caption{Scatter plots of group means by gender and age. For each combination of a category of age, a category of race/ethnicity, and a category of education, male and female define separate subpopulations in the 4-variable cross-classification, with the male group posterior median estimate as the x-coordinate and the female group posterior median estimate as the y-coordinate. The plotting symbols 1 to 4 represent the group's corresponding age category (1: 20 {\textendash} 39, 2: 40 {\textendash} 59, 3: 60 {\textendash} 79, 4: 80+). The gray line is the reference line with slope 1. } \label{fig:mean_gender_age}
\end{figure}

There are multiple ways to present the fitted mean, variance, and correlation estimates of the four health measurements for all of the subgroups categorized by the four demographic characteristics. Here we suggest one graphic display that allows us to examine the relation of the posterior median estimates to a pair of demographic variables.  For a scatter plot, we associate one category of the first variable with the horizontal axis and a second category with the vertical axis.  A digit corresponding to the category of the second variable serves as the plotting symbol.  As an example, Figure \ref{fig:mean_gender_age} illustrates this basic structure with gender as the first variable and age as the second variable, with categories numbered 1 to 4.  For each combination of a category of age, a category of race/ethnicity, and a category of education, male and female define separate subpopulations in the 4-variable cross-classification; the coordinates of the plotted point are the corresponding posterior median estimates.  

\begin{figure}[!ht]
\begin{center}
\includegraphics[height=3in]{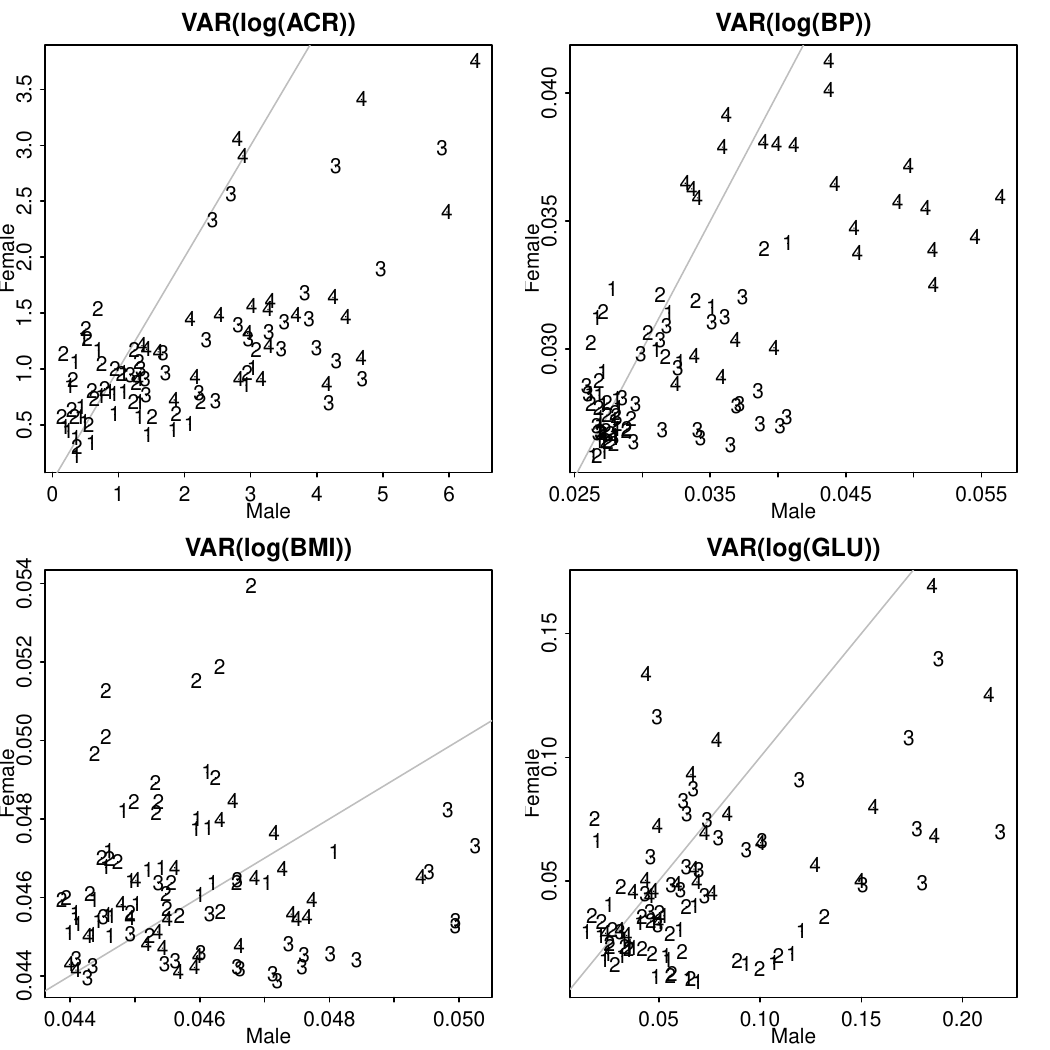}
\end{center}
\caption{ Scatter plots of group variances by gender and age. For each combination of a category of age, a category of race/ethnicity, and a category of education, male and female define separate subpopulations in the 4-variable cross-classification, with the male group posterior median estimate as the x-coordinate and the female group posterior median estimate as the y-coordinate. The plotting symbols 1 to 4 represent the group's corresponding age category (1: 20 {\textendash} 39, 2: 40 {\textendash} 59, 3: 60 {\textendash} 79, 4: 80+). The gray line is the reference line with slope 1. } \label{fig:var_gender_age}
\end{figure}

\begin{figure}[!ht]
\begin{center}
\includegraphics[height=3.5in]{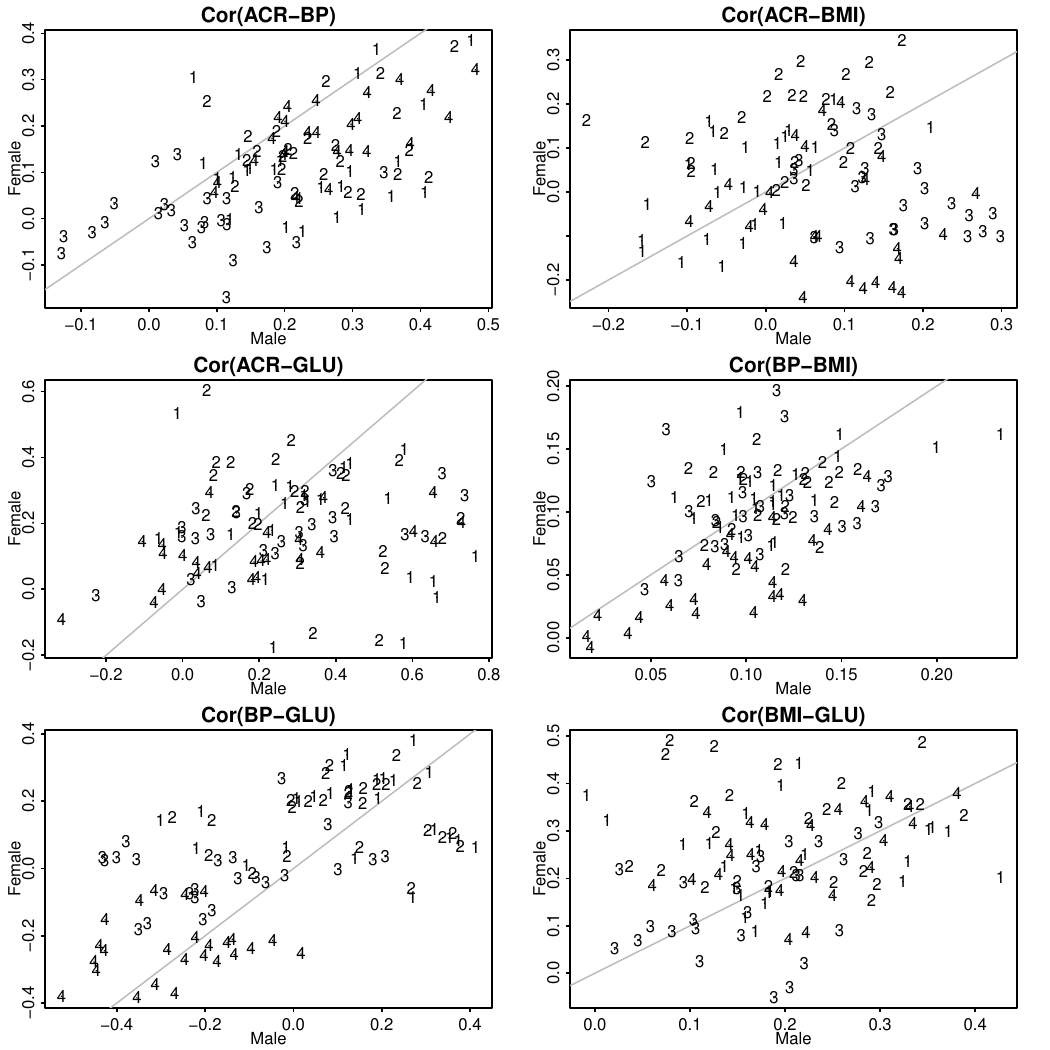}
\end{center}
\caption{Scatter plots of group correlations by gender and age. For each combination of a category of age, a category of race/ethnicity, and a category of education, male and female define separate subpopulations in the 4-variable cross-classification, with the male group posterior median estimate as the x-coordinate and the female group posterior median estimate as the y-coordinate. The plotting symbols 1 to 4 represent the group's corresponding age category (1: 20 {\textendash} 39, 2: 40 {\textendash} 59, 3: 60 {\textendash} 79, 4: 80+). The gray line is the reference line with slope 1. } \label{fig:cor_gender_age}
\end{figure}

These plots show how the mean, variance, and correlation vary with the demographic variables. We highlight a few interesting patterns. Figure \ref{fig:mean_gender_age} shows that females' ACR values are on average higher than the corresponding male's values groups in the younger age groups, but in the older age groups (some of the 60 {\textendash} 79 groups and all of the 80+ groups) males' values are higher. Male groups' blood pressures are almost all higher than the values of the corresponding female groups. The 40 {\textendash} 59 age groups have the highest average blood pressure. More male groups have higher glucose level than the corresponding female groups, and the glucose level seems to increase with age. 

In Figure \ref{fig:var_gender_age}, the variance of ACR and the variance of GLU vary greatly among subpopulations. The ratio of the largest to the smallest posterior median estimate of the standard deviation is 6.76 for ACR and 4.84 for GLU. On the other hand, the variance of BP and the variance of BMI do not vary too much; the ratios are 1.48 and 1.11, respectively. For ACR, most of the male group variance is larger than the  corresponding female group, except for a few younger age groups. Variance generally seems to increase with age. 

In the correlation plot of ACR and BMI (Figure \ref{fig:cor_gender_age}), 60 {\textendash} 79 year-old males have positive correlations, but the corresponding female groups have negative correlations. 40 {\textendash} 59 year-old females have higher correlations than the corresponding male groups. We include similar plots of age and race/ethnicity and age and education in Supplemental Material D (\cite{niu18}). In Supplemental Material E (\cite{niu18}) we present an alternative way of summarizing the results including estimation uncertainties. 

The findings from our model provide some evidence that the mean, variance, and correlation of ACR, BP, BMI, and GLU vary among subpopulations. Therefore, it might not be appropriate to assume a common mean and variance for all subpopulations when deriving reference ranges. More data need to be collected for some subpopulations to determine whether the reference ranges need to be refined.

\section{Discussion}
We use the NHANES data to study how the mean and covariance structure of four health measurements vary among subpopulations. To our knowledge, this is the first attempt to systematically examine how the variances and correlations of those health outcomes vary among subpopulations. We extend the covariance regression model proposed by \cite{hoff_2012} to allow multiple categorical predictors, and we discuss practical issues in fitting complicated datasets with multiple categorical predictors. We select four highly relevant demographic and socio-economic factors to classify the population into subgroups and use the covariance regression model to estimate the mean and covariance parsimoniously for all subpopulations. We discuss guidelines for model selection and evaluation using standard criteria such as AIC for the mean model in conjunction with posterior predictive goodness-of-fit plots for the covariance model. The means, variances, and correlations of those health outcomes all vary among subgroups. The fitted results confirm that the population is heterogeneous and that assuming a single mean and a single covariance for the entire population is not appropriate. We highlight some of the findings that might be of scientific interest. The covariance regression model helps identify subpopulations for which more data might be collected to estimate a separate reference range. In Supplemental Material F (\cite{niu18}) we interpret some of the coefficient estimates. 

We further validate the estimates by comparing the model-based intervals with the sample-based intervals for large groups (41 groups with sample size $>20$). For the variance estimates, the percentage of times those two intervals overlap ranges from 85\% to 95\%, compared with 41\% to 83\% for the homogeneous model. The correlation estimates overlap 98\% to 100\%, compared with 85\% to 95\% for the homogeneous model. The complete set of plots comparing the two sets of intervals is in Supplemental Material G (\cite{niu18}). To consider model mis-specification, the sensitivity analysis (in Supplemental Material H (\cite{niu18})) gives some confidence that the covariance regression model is reasonable and provides reliable estimates.

The model-selection procedure can also allow the search to back up. If, after adding multiple interaction terms, some groups have over-fit, one can remove an interaction and refit, as in stepwise selection. The current selection and evaluation procedure is data-driven and subjective. To develop a systematic model-selection scheme that tries to find the overall best model, one possible approach could explicitly formulate a prior distribution to shrink some of the coefficients toward zero, similar to the idea in \cite{gaskins_2013}.

NHANES uses a complex survey design to select samples that are representative of the U.S. non-institutionalized civilian population. Adjusting the difference between sample and population is very important to obtain unbiased estimates of population quantities. As discussed in \cite{gelman07}, weighting and regression modeling are the two standard ways to accomplish this task. \cite{winship94} compare weighted and unweighted least-squares estimators for linear regression models and conclude that, if the weights depend on only predictors that are included in the model, and the model is true, (unweighted) OLS estimates are unbiased and consistent. In reality, if possible, accounting for the sampling weights is more accurate than using OLS estimates, because we never know whether the regression model is true, and often we cannot include all weight-determining factors and their interactions in the model (\cite{gelman07}). However, incorporating the sampling weights directly can be very difficult for nonstandard models. We therefore choose the regression approach by including the key factors that determine the weights in both the mean and covariance models. NHANES 2009 {\textendash} 2010 oversampled specific age and race/ethnicity groups, as well as pregnant women (\cite{nhanes_guide}). Therefore, the key factors that determine the sampling weights are gender, age, and race/ethnicity, all of which we have included in the proposed joint model. 

To further check for potential biases in our modelings, we compare the model-based estimates of the marginal cell means (such as mean ACR for all males) with the Horvitz-Thomson estimates of marginal cell means. We plot the H-T estimates, 95\% confidence intervals for the H-T estimates, and model-based estimates in Supplemental Material I (\cite{niu18}). The biggest discrepancy lies in the ACR estimates for males and females. This might be due to the fact that during pregnancy the ACR level might change, and we are not able to fully adjust for the oversampling of one gender (female) over the other (male). For most of the other groups, the two estimates are very close, and the model-based estimates almost all fall within the confidence intervals of the H-T estimates. This result gives us some confidence that our model provides approximately unbiased estimates without directly incorporating the sampling weights. It remains an interesting and challenging problem to directly incorporate the weights, thereby fully adjusting for the difference between sample and population.  

In addition to the household survey, NHANES also selected certain participants for physical examinations, based on their demographic and health information. An even smaller proportion had blood tests. Of the 10,537 participants in the 2009 {\textendash} 2010 survey, the numbers who had  blood pressure measurements, urine tests, and body mass measurements range from 7000 to 9000. However, only 3386 participants had a fasting glucose value (blood test). Therefore, the main reduction in sample size is due to the design of the survey and can be viewed as a similar issue as weighting. On the other hand, individual measurements also have missing values; (for example, selected participants failed to show up for exams). Among those who have a GLU value, the missing proportions for the other three measurements (ACR, BP, BMI) are 1\%, 4\%, and 1\%, respectively. According to the NHANES analysis guidelines (http://www.cdc.gov/nchs/tutorials/NHA -NES/Preparing/CleanRecode/Info1.htm), it is usually acceptable to ignore the missing values if the proportion is under 10\%.  Therefore, after accounting for the design variables as discussed above, we assume the missingness due to non-response is ignorable, and we use the complete data for the analysis without any imputation.

The current model assumes multivariate normality of the error term, which is a strong assumption. We assessed some residual plots and did not see serious violations. In essence, we are trying to model the first- and second- order moments that do not rely on normality of the data. The model could be extended to non-normally distributed variables through the generalized linear model framework, as in \cite{pourahmadi_1999}. Another possibility is via semi-parametric copula models, as proposed by \cite{hoff07}. 

In this study, we focus on four specific health outcomes. The method is general enough to be applied to a wide variety of multivariate outcomes.

\section*{Acknowledgements}
The authors thank the reviewers, the associate editor, and the editor for their helpful comments. We are especially grateful for the associate editor's careful review and edits that lead to a stronger article.

\begin{supplement}
\sname{Supplement}\label{supp}
%\stitle{Supplemental Material}
\slink[url]{http://www.e-publications.org/ims/support/dowload/imsart-ims.zip}
\sdatatype{Supp.pdf}
\sdescription{Additional results, tables, and plots mentioned in the text are in the Supplemental Material.}
\end{supplement}

\bibliographystyle{imsart-nameyear}
\bibliography{ref}

\begin{thebibliography}{31}
% BibTex style file: imsart-nameyear.bst, 2017-11-03
% Default style options (sort=1,type=nameyear).
% Used options (sort=1,type=nameyear).

\bibitem[\protect\citeauthoryear{Akaike}{1973}]{aic73}
\begin{binproceedings}[author]
\bauthor{\bsnm{Akaike},~\bfnm{H.}\binits{H.}}
(\byear{1973}).
\btitle{Information Theory and an Extension of the Maximum Likelihood
  Principle}.
In \bbooktitle{B. N. Petrov, and F. Csaki (Eds.), Second International
  Symposium on Information Theory}
\bpages{267--281}.
\end{binproceedings}
\endbibitem

\bibitem[\protect\citeauthoryear{Boik}{2002}]{boik_2002}
\begin{barticle}[author]
\bauthor{\bsnm{Boik},~\bfnm{R.~J.}\binits{R.~J.}}
(\byear{2002}).
\btitle{Spectral models for covariance matrices}.
\bjournal{Biometrika}
\bvolume{89}
\bpages{159--182}.
\end{barticle}
\endbibitem

\bibitem[\protect\citeauthoryear{Boik}{2003}]{boik_2003}
\begin{barticle}[author]
\bauthor{\bsnm{Boik},~\bfnm{R.~J.}\binits{R.~J.}}
(\byear{2003}).
\btitle{Principal component models for correlation matrices}.
\bjournal{Biometrika}
\bvolume{90}
\bpages{679--701}.
\end{barticle}
\endbibitem

\bibitem[\protect\citeauthoryear{{CDC/NCHS}}{2010a}]{nhanes_2010}
\begin{bmanual}[author]
\bauthor{\bsnm{{CDC/NCHS}}}
(\byear{2010}a).
\btitle{National Health and Nutrition Examination Survey Data, 2009-2010}
\bpublisher{U.S. Department of Health and Human Services, Centers for Disease
  Control and Prevention, National Center for Health Statistics},
\baddress{Hyattsville, MD}.
\end{bmanual}
\endbibitem

\bibitem[\protect\citeauthoryear{{CDC/NCHS}}{2010b}]{nhanes_guide}
\begin{bmanual}[author]
\bauthor{\bsnm{{CDC/NCHS}}}
(\byear{2010}b).
\btitle{National Health and Nutrition Examination Survey: Analytic Guidelines,
  1999-2010}
\bpublisher{U.S. Department of Health and Human Services, Centers for Disease
  Control and Prevention, National Center for Health Statistics},
\baddress{Hyattsville, MD}.
\end{bmanual}
\endbibitem

\bibitem[\protect\citeauthoryear{Chiu, Leonard and Tsui}{1996}]{chiu_1996}
\begin{barticle}[author]
\bauthor{\bsnm{Chiu},~\bfnm{Tom Y.~M.}\binits{T.~Y.~M.}},
  \bauthor{\bsnm{Leonard},~\bfnm{Tom}\binits{T.}} \AND
  \bauthor{\bsnm{Tsui},~\bfnm{Kam-Wah}\binits{K.-W.}}
(\byear{1996}).
\btitle{The matrix-logarithmic covariance model}.
\bjournal{J. Amer. Statist. Assoc.}
\bvolume{91}
\bpages{198--210}.
\bmrnumber{MR1394074 (97a:62120)}
\end{barticle}
\endbibitem

\bibitem[\protect\citeauthoryear{CLSI}{2008}]{CLSI2008}
\begin{bbook}[author]
\bauthor{\bsnm{CLSI}}
(\byear{2008}).
\btitle{Defining, Establishing, and Verifying Reference Intervals in the
  Clinical Laboratory: Approved Guideline},
\bedition{3rd} ed.
\bpublisher{CLSI document EP28-A3c. Wayne, PA: Clinical and Laboratory
  Standards Institute}.
\end{bbook}
\endbibitem

\bibitem[\protect\citeauthoryear{Cox and Reid}{1987}]{cox_1987}
\begin{barticle}[author]
\bauthor{\bsnm{Cox},~\bfnm{D.~R.}\binits{D.~R.}} \AND
  \bauthor{\bsnm{Reid},~\bfnm{N.}\binits{N.}}
(\byear{1987}).
\btitle{Parameter orthogonality and approximate conditional inference (with
  discussion)}.
\bjournal{Journal of the Royal Statistical Society, Series B}
\bvolume{49}
\bpages{1--39}.
\end{barticle}
\endbibitem

\bibitem[\protect\citeauthoryear{Cripps, Carter and Kohn}{2005}]{cripps_2005}
\begin{bincollection}[author]
\bauthor{\bsnm{Cripps},~\bfnm{Edward}\binits{E.}},
  \bauthor{\bsnm{Carter},~\bfnm{Chris}\binits{C.}} \AND
  \bauthor{\bsnm{Kohn},~\bfnm{Robert}\binits{R.}}
(\byear{2005}).
\btitle{Variable selection and Covariance selection in multivariate Regression
  Models}.
In \bbooktitle{Handbook of Statistics 25: Bayesian Thinking: Modeling and
  Computation}
(\beditor{\bfnm{Dipak}\binits{D.}~\bsnm{Dey}} \AND
  \beditor{\bfnm{C.~R.}\binits{C.~R.}~\bsnm{Rao}}, eds.)
\bpages{519–-552}.
\bpublisher{Amsterdam: North-Holland}.
\end{bincollection}
\endbibitem

\bibitem[\protect\citeauthoryear{Engle and Kroner}{1995}]{engle_kroner_1995}
\begin{barticle}[author]
\bauthor{\bsnm{Engle},~\bfnm{Robert~F.}\binits{R.~F.}} \AND
  \bauthor{\bsnm{Kroner},~\bfnm{Kenneth~F.}\binits{K.~F.}}
(\byear{1995}).
\btitle{Multivariate simultaneous generalized {ARCH}}.
\bjournal{Econometric Theory}
\bvolume{11}
\bpages{122--150}.
\bdoi{10.1017/S0266466600009063}
\bmrnumber{MR1325104 (96d:62215)}
\end{barticle}
\endbibitem

\bibitem[\protect\citeauthoryear{Fong, Li and An}{2006}]{fong_2006}
\begin{barticle}[author]
\bauthor{\bsnm{Fong},~\bfnm{P.~W.}\binits{P.~W.}},
  \bauthor{\bsnm{Li},~\bfnm{W.~K.}\binits{W.~K.}} \AND
  \bauthor{\bsnm{An},~\bfnm{Hong-Zhi}\binits{H.-Z.}}
(\byear{2006}).
\btitle{A simple multivariate {ARCH} model specified by random coefficients}.
\bjournal{Comput. Statist. Data Anal.}
\bvolume{51}
\bpages{1779--1802}.
\bdoi{10.1016/j.csda.2005.11.019}
\bmrnumber{MR2307543}
\end{barticle}
\endbibitem

\bibitem[\protect\citeauthoryear{Foulds, Bredin and Warburton}{2012}]{foulds12}
\begin{barticle}[author]
\bauthor{\bsnm{Foulds},~\bfnm{HJA}\binits{H.}},
  \bauthor{\bsnm{Bredin},~\bfnm{SSD}\binits{S.}} \AND
  \bauthor{\bsnm{Warburton},~\bfnm{DER}\binits{D.}}
(\byear{2012}).
\btitle{The Relationship between Diabetes and Obesity across Different
  Ethnicities}.
\bjournal{Journal of Diabetes and Metabolism}
\bvolume{3}.
\bdoi{10.4172/2155-6156.1000228}
\end{barticle}
\endbibitem

\bibitem[\protect\citeauthoryear{Fraser et~al.}{2012}]{fraser_2012}
\begin{barticle}[author]
\bauthor{\bsnm{Fraser},~\bfnm{Simon D.~S.}\binits{S.~D.~S.}},
  \bauthor{\bsnm{Roderick},~\bfnm{Paul~J.}\binits{P.~J.}},
  \bauthor{\bsnm{Mclntyre},~\bfnm{Natasha~J.}\binits{N.~J.}},
  \bauthor{\bsnm{Harris},~\bfnm{Scott}\binits{S.}},
  \bauthor{\bsnm{Mclntyre},~\bfnm{Christopher~W.}\binits{C.~W.}},
  \bauthor{\bsnm{Fluck},~\bfnm{Richard~J.}\binits{R.~J.}} \AND
  \bauthor{\bsnm{Taal},~\bfnm{Maarten~W.}\binits{M.~W.}}
(\byear{2012}).
\btitle{Socio-Economic Disparities in the Distribution of Cardiovascular Risk
  in Chronic Kidney Disease Stage 3}.
\bjournal{Nephron Clinical Practice}
\bvolume{122}
\bpages{58--65}.
\end{barticle}
\endbibitem

\bibitem[\protect\citeauthoryear{Gaskins and Daniels}{2013}]{gaskins_2013}
\begin{barticle}[author]
\bauthor{\bsnm{Gaskins},~\bfnm{Jeremy~T.}\binits{J.~T.}} \AND
  \bauthor{\bsnm{Daniels},~\bfnm{Michael~J.}\binits{M.~J.}}
(\byear{2013}).
\btitle{A nonparametric prior for simultaneous covariance estimation}.
\bjournal{Biometrika}
\bvolume{100}
\bpages{125--138}.
\end{barticle}
\endbibitem

\bibitem[\protect\citeauthoryear{Gelman}{2007}]{gelman07}
\begin{barticle}[author]
\bauthor{\bsnm{Gelman},~\bfnm{Andrew}\binits{A.}}
(\byear{2007}).
\btitle{Struggles with Survey Weighting and Regression Modeling}.
\bjournal{Statistical Science}
\bvolume{22}
\bpages{153--164}.
\end{barticle}
\endbibitem

\bibitem[\protect\citeauthoryear{Guttman}{1967}]{guttman67}
\begin{barticle}[author]
\bauthor{\bsnm{Guttman},~\bfnm{I.}\binits{I.}}
(\byear{1967}).
\btitle{The use of the concept of a future observation in goodness-of-fit
  problems}.
\bjournal{Journal of the Royal Statistical Society: Series B}
\bvolume{29}
\bpages{83--100}.
\end{barticle}
\endbibitem

\bibitem[\protect\citeauthoryear{Harris and Boyd}{1990}]{harris90}
\begin{barticle}[author]
\bauthor{\bsnm{Harris},~\bfnm{E~K}\binits{E.~K.}} \AND
  \bauthor{\bsnm{Boyd},~\bfnm{J~C}\binits{J.~C.}}
(\byear{1990}).
\btitle{On dividing reference data into subgroups to produce separate reference
  ranges.}
\bjournal{Clinical Chemistry}
\bvolume{36}
\bpages{265--270}.
\end{barticle}
\endbibitem

\bibitem[\protect\citeauthoryear{Hoff}{2007}]{hoff07}
\begin{barticle}[author]
\bauthor{\bsnm{Hoff},~\bfnm{Peter~D.}\binits{P.~D.}}
(\byear{2007}).
\btitle{Extending the rank likelihood for semiparametric copula estimation}.
\bjournal{Annals of Applied Statistics}
\bvolume{1}
\bpages{265--283}.
\end{barticle}
\endbibitem

\bibitem[\protect\citeauthoryear{Hoff}{2009}]{hoff_2009}
\begin{barticle}[author]
\bauthor{\bsnm{Hoff},~\bfnm{Peter~D.}\binits{P.~D.}}
(\byear{2009}).
\btitle{A hierarchical eigenmodel for pooled covariance estimation}.
\bjournal{Journal of the Royal Statistical Society, Series B}
\bvolume{71}
\bpages{971--992}.
\end{barticle}
\endbibitem

\bibitem[\protect\citeauthoryear{Hoff and Niu}{2012}]{hoff_2012}
\begin{barticle}[author]
\bauthor{\bsnm{Hoff},~\bfnm{Peter~D.}\binits{P.~D.}} \AND
  \bauthor{\bsnm{Niu},~\bfnm{Xiaoyue}\binits{X.}}
(\byear{2012}).
\btitle{A Covariance Regression Model}.
\bjournal{Statistica Sinica}
\bvolume{22}
\bpages{729--753}.
\end{barticle}
\endbibitem

\bibitem[\protect\citeauthoryear{KDIGO}{2013}]{kdigo12}
\begin{barticle}[author]
\bauthor{\bsnm{KDIGO}}
(\byear{2013}).
\btitle{Kidney {D}isease: Improving {G}lobal {O}utcomes ({KDIGO}) {CKD} {W}ork
  {G}roup. {KDIGO} 2012 Clinical Practice Guideline for the Evaluation and
  Management of Chronic Kidney Disease}.
\bjournal{Kidney {I}nter., Suppl.}
\bvolume{3}
\bpages{1--150}.
\end{barticle}
\endbibitem

\bibitem[\protect\citeauthoryear{Liang and Zeger}{1986}]{liang_1986}
\begin{barticle}[author]
\bauthor{\bsnm{Liang},~\bfnm{Kung-Yee}\binits{K.-Y.}} \AND
  \bauthor{\bsnm{Zeger},~\bfnm{Scott~L.}\binits{S.~L.}}
(\byear{1986}).
\btitle{Longitudinal data analysis using generalized linear models}.
\bjournal{Biometrika}
\bvolume{73}
\bpages{13--22}.
\end{barticle}
\endbibitem

\bibitem[\protect\citeauthoryear{Mattix et~al.}{2002}]{mattix02}
\begin{barticle}[author]
\bauthor{\bsnm{Mattix},~\bfnm{Holly~J.}\binits{H.~J.}},
  \bauthor{\bsnm{Hsu},~\bfnm{Chi-Yuan}\binits{C.-Y.}},
  \bauthor{\bsnm{Shaykevich},~\bfnm{Shimon}\binits{S.}} \AND
  \bauthor{\bsnm{Curhan},~\bfnm{Gary}\binits{G.}}
(\byear{2002}).
\btitle{Use of the Albumin/Creatinine Ratio to Detect Microalbuminuria:
  Implications of Sex and Race}.
\bjournal{Journal of the American Society of Nephrology}
\bvolume{13}
\bpages{1034--1039}.
\end{barticle}
\endbibitem

\bibitem[\protect\citeauthoryear{McCullagh and Nelder}{1989}]{glm}
\begin{bbook}[author]
\bauthor{\bsnm{McCullagh},~\bfnm{P.}\binits{P.}} \AND
  \bauthor{\bsnm{Nelder},~\bfnm{John~A.}\binits{J.~A.}}
(\byear{1989}).
\btitle{Generalized Linear Models, 2nd ed}.
\bpublisher{Chapman \& Hall/CRC}, \baddress{London}.
\end{bbook}
\endbibitem

\bibitem[\protect\citeauthoryear{NIDDK}{2013}]{ckd_prev}
\begin{bmanual}[author]
\bauthor{\bsnm{NIDDK}}
(\byear{2013}).
\btitle{U.{S}. Renal Data System, {USRDS} 2013 Annual Data Report: Atlas of
  Chronic Kidney Disease and End-Stage Renal Disease in the {U}nited {S}tates}
\bpublisher{National Institutes of Health, National Institute of Diabetes and
  Digestive and Kidney Diseases},
\baddress{Bethesda, MD}.
\end{bmanual}
\endbibitem

\bibitem[\protect\citeauthoryear{Niu and Hoff}{2018}]{niu18}
\begin{barticle}[author]
\bauthor{\bsnm{Niu},~\bfnm{Xiaoyue}\binits{X.}} \AND
  \bauthor{\bsnm{Hoff},~\bfnm{Peter~D.}\binits{P.~D.}}
(\byear{2018}).
\btitle{Supplement to ``Joint Mean and Covariance Modeling of Multiple Health
  Outcome Measures''}.
\end{barticle}
\endbibitem

\bibitem[\protect\citeauthoryear{Pourahmadi}{1999}]{pourahmadi_1999}
\begin{barticle}[author]
\bauthor{\bsnm{Pourahmadi},~\bfnm{Mohsen}\binits{M.}}
(\byear{1999}).
\btitle{Joint mean-covariance models with applications to longitudinal data:
  unconstrained parameterisation}.
\bjournal{Biometrika}
\bvolume{86}
\bpages{677--690}.
\bdoi{10.1093/biomet/86.3.677}
\bmrnumber{MR1723786 (2000g:62178)}
\end{barticle}
\endbibitem

\bibitem[\protect\citeauthoryear{Pourahmadi}{2011}]{pourahmadi_2011}
\begin{barticle}[author]
\bauthor{\bsnm{Pourahmadi},~\bfnm{Mohsen}\binits{M.}}
(\byear{2011}).
\btitle{Covariance Estimation: The {GLM} and Regularization Perspectives}.
\bjournal{Statistical Science}
\bvolume{26}
\bpages{369--387}.
\end{barticle}
\endbibitem

\bibitem[\protect\citeauthoryear{Rubin}{1984}]{rubin84}
\begin{barticle}[author]
\bauthor{\bsnm{Rubin},~\bfnm{Don~B.}\binits{D.~B.}}
(\byear{1984}).
\btitle{Bayesian justifiable and relevant frequency calculations for the
  applied statistician}.
\bjournal{Annals of Statistics}
\bvolume{12}
\bpages{1151--1172}.
\end{barticle}
\endbibitem

\bibitem[\protect\citeauthoryear{Winship and Radbill}{1994}]{winship94}
\begin{barticle}[author]
\bauthor{\bsnm{Winship},~\bfnm{Christopher}\binits{C.}} \AND
  \bauthor{\bsnm{Radbill},~\bfnm{Larry}\binits{L.}}
(\byear{1994}).
\btitle{Sampling Weights and Regression Analysis}.
\bjournal{Sociological Methods and Research}
\bvolume{23}
\bpages{230--257}.
\end{barticle}
\endbibitem

\bibitem[\protect\citeauthoryear{Zeger and Liang}{1986}]{zeger_1986}
\begin{barticle}[author]
\bauthor{\bsnm{Zeger},~\bfnm{Scott~L.}\binits{S.~L.}} \AND
  \bauthor{\bsnm{Liang},~\bfnm{Kung-Yee}\binits{K.-Y.}}
(\byear{1986}).
\btitle{Longitudinal data analysis for discrete and continuous outcomes}.
\bjournal{Biometrics}
\bvolume{42}
\bpages{121--130}.
\end{barticle}
\endbibitem

\end{thebibliography}

\end{document}